\documentstyle[12pt]{article}

\begin{document}
\date{}
\title{EQUILIBRIUM CONFIGURATION OF BLACK HOLES AND THE INVERSE
SCATTERING METHOD\footnote{journal reference:Theor. Math. Phys.
v.111, p.667(1997)}}
\author{G. G. Varzugin}
\maketitle
\begin{center}{\it
Laboratory of Complex System Theory, Physics Institute of
St-Petersburg University, St. Petersburg, Peterhof, Ulyanovskaya
1, 198904. E-mail varzugin@paloma.spbu.ru}\end{center} \vskip1cm
\begin{abstract}
The inverse scattering method is applied to the investigation of
the equilibrium configuration of black holes. A study of the boundary
problem corresponding to this configuration shows that
any axially symmetric, stationary solution of the
Einstein equations with disconnected event horizon must
belong to the class of Belinskii-Zakharov solutions.
Relationships between the angular momenta and angular
velocities of black holes are derived.
\end{abstract}

\section{Introduction}
In the present paper, we study solutions to the Einstein equations in
a vacuum that are stationary, axially symmetric, and asymptotically
flat. The main purpose is to elaborate a description in which two
classes of ideas, black hole theory and the theory of completely
integrable equations, are unified.

In black hole physics, the well-known result claims that "black holes
have no hair". In other words, each black hole is uniquely determined
by its mass and angular momentum and is nothing but the Kerr
solution. Many authors contributed to this result and we do not have
room to give the exhaustive references. For our purpose, only the
final stage of the proof is important, namely, Carter's
classification of all axially symmetric solutions having connected
horizon \cite{Carter75}. Carter demonstrated that these solutions
must solve the boundary problem for a system of elliptic nonlinear
equations, the Kerr solution being one of the possible solution to
this boundary problem. The uniqueness of this solution was proven in
\cite{Robinson75}.

Can one do without requiring the connectedness of the horizon? In
other words, do solutions exists that correspond to an equilibrium
configuration of black holes? To the best of the author knowledge,
these questions still have no exhaustive answers (though some results
can be found in \cite{Weinstein94,Li91}). Also, these questions can
be formulated in terms of the boundary problem from \cite{Carter75}
because, there, the connectedness of the horizon played no role. This
boundary problem is formulated in the following section.

It is well known that the Einstein equations with two commuting
symmetries belong to a wide class of systems that can be
integrated by the methods of the inverse scattering problem. This
was shown in \cite{Belinskii-Zakharov78} by Belinskii and
Zakharov, who also investigated the axially symmetric case for
which the $2N$-soliton solution was constructed
\cite{Belinskii-Zakharov79}. This solution was interpreted as the
solution corresponding to the $N$ Kerr-NUT black holes. As we show
in the present paper, the $2N$-soliton solution of
Belinskii-Zakharov indeed contains all possible solutions (if any
exist) corresponding to an equilibrium configuration of rotating
black holes. Therefore, we reduce the question of the existence of
solutions with disconnected horizon to the investigation of some
subclass of the soliton solutions. This subclass is parameterized
by the distances between black holes, the angular momenta, and the
masses of the black holes. However, in general case, solutions
from this have a conical singularity on the symmetry axis, which
hinders the existence of solutions with disconnected horizon. It
is still unclear whether it possible to choose the parameters in a
way that removes this conical singularity. Most likely, the answer
is negative \cite{Weinstein94,Li91}.

The presence of the conical singularity does not make these solutions
physically meaningless. On the contrary, the conical singularity
itself has the physical sense of the interaction force between the
black holes. This interpretation was proposed by Weyl; for details
see \cite{Weinstein90}.

\section{Boundary conditions}

In this section, we present some basic facts about the stationary
axially symmetric solutions to the Einstein equations and formulate
the boundary problem corresponding to an equilibrium configuration
for black holes. The results of \cite{Carter75} are crucial for us
and we refer the reader there for details.

Recall that Lorentz manifold is stationary and axially symmetric
if it possesses two commuting one-parameter isometry groups that
are isomorphic to $R$ and $SO(2)$ respectively. In other words,
there exist two Killing vector fields $k^a$ and $m^a$, which
commute with each other, such that the vector $k^a$ is time-like
and the vector $m^a$ is space-like. For the stationary and axially
symmetric vacuum (or electrovacuum) space-time one can always
choose the coordinate system in which the metric acquires the
following form: $$ds^2=-Vdt^2+2Wdtd\phi+Xd\phi^2+{X\over\rho^2}
e^{\beta}(d\rho^2+dz^2).$$ Here the metric coefficients depend
only on $\rho$ and $z$, and $k^a\partial/\partial
x^a=\partial/\partial t$, $m^a\partial/\partial
x^a=\partial/\partial\phi$.  Then $V,W,X,\rho$ have a geometric
meaning, $$V=-k^ak_a,\; X=m^am_a,\; W=k^am_a, \eqno(2.1a ) $$
$$VX+W^2=\rho^2=-\rho_{ab}\rho^{ab}\; (\rho_{ab}=2k_{[a}m_{b]}).
\eqno(2.1b)$$ The multitudes of the Killing vectors are normalized
as follows: $V\rightarrow1$ at infinity and ${X^{,a}X_{,a}\over
4X}\rightarrow1$ on the symmetry axis. The coordinates $\phi,\rho$
and $z$ become cylindrical (Weyl) coordinates.

The set of points with $\rho=0, X=0$ is the symmetry axis, while
the set of points with $\rho=0, X>0$ is the event horizon. Let the
event horizon have $N$ connected components and $l^a$ be an
isotropic vector that is orthogonal to the event horizon. Then, in
each connected component of the horizon, we can choose such a
normalization of $l^a$ that $$l^a=k^a+\Omega_im^a,$$ where
$\Omega_i$ is some constant whose physical meaning is the angular
velocity of the black hole.Let $z_1,z_2,\ldots,z_{2N}$ be the
points of intersection of the horizon and the symmetry axis
enumerated in increasing order.

We pass to a new coordinate system,
$$\rho^2=(\lambda^2-m_i^2)(1-\mu^2),
m_i={z_{2i}-z_{2i-1}\over2},$$
$$z-{z_{2i}+z_{2i-1}\over2}=\lambda\mu.$$ In this coordinate
system, the necessary conditions of regularity of the symmetry
axis and the horizon are formulated as follows:
$$X(\lambda,\mu)=(1-\mu^2)\hat X(\lambda,\mu)$$
$$W^\dagger(\lambda,\mu)=(\lambda^2-m_i^2)(1-\mu^2)\hat
W(\lambda,\mu)\eqno(2.2)$$
$$V^\dagger(\lambda,\mu)=(\lambda^2-m_i^2)\hat V(\lambda,\mu) $$
Here $\hat X,\hat W,\hat V$ are smooth functions nowhere equal to
zero and $$V^\dagger=-l^al_a,\; W^\dagger=l^am_a.$$ From (2.2) one
can easily obtain the boundary conditions for the first group of
Einstein equations, $$d\ast\rho dgg^{-1}=0,\; g=\pmatrix {-V&W\cr
W&X\cr },\eqno(2.3)$$ where $\ast$ is the Hodge operator $\ast
d\rho=dz, \ast dz=-d\rho$. Indeed using (2.2), one can easily
prove that $$\rho g_{,\rho}g^{-1}=\pmatrix{0&O(1)\cr0&2\cr},\;
\rho\rightarrow0,\;z\in\Gamma, \eqno(2.4a)$$ $$\hat\Omega_i\rho
g_{,\rho}g^{-1} \hat\Omega_i^{-1}=\pmatrix{2&0\cr
O(1)&0\cr},\;\rho\rightarrow0 \;z\in I_i,\;
\hat\Omega_i=\pmatrix{1&\Omega_i\cr0&1}.\eqno(2.4b)$$ $$\rho
g_{,z}g^{-1}=O(1),\;\rho\rightarrow0,\; z\in R.\eqno(2.4c)$$ Here
$\Gamma$ is the symmetry axis consisting of $N+1$ connected
components, $$\Gamma=\bigcup\Gamma_j=R\setminus\bigcup I_i,\;\;\;
j=1,\ldots,N+1,\;\;\; i=1,\ldots,N,\;\;\; I_i=(z_{2i-1},z_{2i}).$$
The symbol $O(1)$ denotes a uniformly bounded function on
corresponding interval. It follows from (2.2) that (2.4c) tends to
zero almost everywhere except the points $z_k$; however for our
purposes, a uniform boundness suffices. The function $g(z,\rho)$
is taken to be smooth at all points except the points $(z_k,0)$.

As we demonstrated below, Eqs. (2.4) completely determine the
solution to Eqs. (2.3). At the same time $\Omega_i$ and $z_i$
are the independent parameters of the boundary problem; these
parameters can be choose arbitrary. The behavior of
$g$ at infinity is discussed at the end of this section.

An alternative approach, in which the main parameters are angular
momenta rather than angular velocities, exists
\cite{Weinstein90,Weinstein92}. To show this let us introduce the
Ernst potentials $$\rho\ast dgg^{-1}=\pmatrix{-dY^{12}&d\tilde
Y\cr -dY&dY^{21}\cr},\;\;d(Y^{21}-Y^{12})=2dz.\eqno(2.5)$$ Here
$Y$ is the Ernst potential that is determined by the space-like
Killing vector field, while $\tilde Y$ is the Ernst potential
determined by the time-like Killing vector field. Then system
(2.3) can be rewritten in equivalent form, $$d({\rho\ast dX\over
X})-{\rho\over X^2}\ast dY\land dY=0,$$ $$d({\rho\ast dY\over
X})+{\rho\over X^2}\ast dX\land dY=0,\eqno(2.6a)$$
$$d\Omega=-{\rho\ast dY\over X^2},\;\;\; W=\Omega X. \eqno(2.6b)$$
From (2.4a) or (2.2) we can see that $$\rho\partial_\rho\ln
X\rightarrow2,\;\rho\rightarrow0, \; z\in\Gamma
;\;Y|_{\Gamma_i}=c_i,\eqno(2.7)$$ where $c_i$ are some constants
that are independent parameters. In
\cite{Weinstein94,Weinstein90,Weinstein92}, it was proved that
(2.6) has a unique solution satisfying (2.7) and some condition at
infinity for all $z_i$ and $c_i$. In fact, the latter condition is
equivalent to the asymptotic flatness of the metric. Note that
(2.4b) follows from (2.6b) provided that $\Omega_i$ is defined as
follows: $$\Omega|_{I_i}=-\Omega_i,\eqno(2.8)$$ Thus, $\Omega_i$
is a function of $z_i$ and $c_i$.

The constants $c_i$ unambiguously determine the angular momentum of
all black holes. Indeed, let us define the angular momenta of a black
hole using the Komar form:
$$L_i={1\over16\pi}\int_{S_i}m^{a;b}dS_{ab},$$
where $S_i$ is a two-surface surrounding the black hole.
Choosing $S_i$ to be the surface of revolution of the curve $C_i$
connecting the components of the axis $\Gamma_{i+1}$ and $\Gamma_i$,
we obtain
$$L_i={1\over8}\int_{C_i}{1\over\rho}\ast(XdW-WdX)=
{1\over8}\int_{C_i}dY={1\over8}(c_{i+1}-c_i).\eqno(2.9)$$

The mass of the black hole can be defined as the following
integral \cite{Carter75}:
$$M_i=-{1\over8\pi}\int_{S_i}k^{a;b}dS_{ab}.$$ Proceeding as for
finding the angular momentum, we obtain that
$$M_i={1\over4}\int_{C_i}{1\over\rho}\ast(XdV+WdW)=
-{1\over4}\int_{C_i}dY^{12}.$$ Contracting the contour $C_i$ to
the horizon and using (2.4b), we find that
$$M_i={1\over4}\int_{I_i}(2+\Omega_i Y_{,z})dz=
m_i+2\Omega_iL_i,\eqno(2.10)$$ where $m_i=(z_{2j}-z_{2j-1})/2$.

At infinity, we impose the following conditions: $$W=\rho^2
O({1\over r^3}),\;\; X=\rho^2(1+O({1\over r})). \eqno(2.11)$$
where $r=\sqrt{\rho^2+z^2}$. Asymptotic formulas (2.11) are
assumed to be differentiable at least twice. Formulas (2.11) mean
that the metric tensor $g$ tends to the Minkowski tensor in
cylindrical coordinates. From (2.11) we obtain
$$g_{,z}g^{-1}=\pmatrix{O(1/r^2)&O(1/r^4)\cr \rho^2
O(1/r^4)&O(1/r^2)\cr},\;\; V=1+O(1/r),\eqno(2.12a)$$ $$\rho
g_{,\rho}g^{-1}-\pmatrix{0&0\cr0&2}=\pmatrix{O(1/r) &O(1/r^3)\cr
\rho^2 O(1/r^3)&O(1/r)\cr}.\eqno(2.12b)$$ When determining the
angular velocities from (2.8), we should normalize $\Omega$ in
accordance with (2.11), i.e. $\Omega=O(1/r^3)$.

The second group of Einstein equations allows one to determine the
coefficient $e^\beta$ from the matrix $g$. Using (2.4) or
(2.7), one can show that
$\partial_z\beta=0$ for$\rho=0$ and $z\in\Gamma$, i.e.
$\beta|_{\Gamma_i}=b_i$, where $b_i$ are some constant.
The conical singularity on the symmetry axis is absent iff
$b_i=0$. However, $b_i$ cannot be treated as independent parameters
since they are functions of $z_k$ and $c_k$. In the preset paper, we
restrict ourselves to the study of boundary problem (2.4), (2.12) and
do not discuss the properties of $b_i$.

\section{Auxiliary linear problem}

System of equations (2.3) is the compatibility condition for the
following pair of matrix linear differential equations
\cite{Belinskii-Zakharov78,Belinskii-Zakharov79}:
$$D_1\psi={\rho^2g_{,z}g^{-1}-\omega\rho g_{,\rho}g^{-1}\over
\omega^2+\rho^2}\psi,\;\; D_2\psi={\rho^2g_{,\rho}g^{-1}+\omega\rho
g_{,z}g^{-1}\over\omega^2+ \rho^2}\psi.\eqno(3.1)$$
Here $D_1$ and $D_2$ are the commuting differential operators:
$$D_1=\partial_z-{2\omega^2\over\omega^2+\rho^2}\partial_\omega,
\;\;D_2=\partial_\rho+{2\omega\rho\over\omega^2+\rho^2}
\partial_\omega,$$
and $\omega$ is a complex parameter that does not depend on the
coordinates. We also use the $U-V$ pair representation in which
$\omega$ is a dependent parameter. To be more precise, let $\omega$
be a root of the equation
$$\omega^2-2\omega(k-z)-\rho^2=0,\eqno(3.2)$$
where $k$, in turn, is independent spectral parameter.
Using (3.2), one can easily check that
$$\partial_z\omega=-{2\omega^2\over\omega^2+\rho^2},\;\;
\partial_\rho\omega={2\omega\rho\over\omega^2+\rho^2}.\eqno(3.3)$$
Passing from $\psi(\omega)$ to $\psi^\prime(k)=\psi(\omega(k))$,
we obtain from (3.1) that $$\partial_z\psi(k)=A(z,\rho
,k)\psi(k),\;\; A={\rho^2 g_{,z}g^{-1}-\omega\rho
g_{,\rho}g^{-1}\over \omega^2+\rho^2}, \eqno(3.4a)$$
$$\partial_\rho\psi(k)=B(z,\rho ,k)\psi(k),\;\;
B={\rho^2g_{,\rho}g^{-1}+\omega\rho g_{,z}g^{-1}
\over\omega^2+\rho^2} \eqno(3.4b)$$ Hereafter, we omit the prime
for brevity. It is worth mentioning that Eqs. (3.4) are equivalent
to Eqs. (3.1) only if we take $\omega$ to be the multivalued
function in (3.4). Fixing the branch of the root in (3.4), we find
the solution to system (3.1) only in the analyticity domain of
$\omega(k)$.

In the present paper, we follow the general scheme for investigating
integrable equations \cite{Takhtajan-Faddeev}.

Since Eq.(3.2) is invariant with respect to the transformation
$\omega \rightarrow -\rho^2/\omega$, we can fix the branch of the
multivalued function $\omega(k)$, stipulating that the inequality
$|\omega|>\rho$ holds. Then, from (3.2), we obtain
$$\omega\rightarrow2(k-z),\;\rho\rightarrow0;\;
\omega\rightarrow2(k-z),\; z\rightarrow\infty;\;
\omega\rightarrow2(k-z),\; k\rightarrow\infty\eqno(3.5)$$ After
choosing the branch of the root, we can introduce the monodromy
matrix $T(z,y)$, which, by definition, is a solution to (3.4a)
such that $T(y,y)=I.$ Note that for $\rho\rightarrow0$,
$$A(z,\rho,k)\rightarrow{1\over2}{1\over z-k}
\pmatrix{0&\partial_z\tilde Y\cr0&2\cr},\;\;z\in\Gamma,$$
$$A(z,\rho,k)\rightarrow{1\over2}{1\over z-k}
\hat\Omega_i^{-1}\pmatrix{2&0\cr -\partial_zY&0\cr}
\hat\Omega_i,\;\;z\in I_i.$$ Here we took into account (3.5) and
boundary condition (2.4). Hence Eq. (3.4a) can be easily
integrated at $\rho=0$. As a result, the explicit formulas for the
monodromy matrix are $$T(z,y)=\pmatrix{1&-{\tilde Y(z)-\tilde
Y(y)\over 2(k-y)}\cr0&{k-z\over k-y}\cr},\;\;
z,y\in\Gamma_k,\eqno(3.7)$$ where $\Gamma_k$ is the connected
component of the symmetry axis and
$$T(z,y)=\hat\Omega_i^{-1}\pmatrix{{k-z\over k-y}&0\cr
{Y(z)-Y(y)\over2(k-y)}&1\cr}\hat\Omega_i,\;\; z,y\in I_i.
\eqno(3.8)$$

Let $e(z,\rho,k)$ be the solution to (3.4) with the Minkowski tensor
in cylindrical coordinates ($V=1, W=0,
X=\rho^2$),
$$e(z,k)=\pmatrix{1&0\cr0&\omega(z,k)\cr},\;
\partial_ze=A_0e,\; A_0=\pmatrix{0&0\cr
0&-{2\omega\over\omega^2+\rho^2}\cr}.\eqno(3.9)$$ Let us define
the Jost functions and reduced monodromy matrix,
$$\Psi^\pm(z,k)=\lim_{y\rightarrow\pm\infty}T(z,y)e(y),
\eqno(3,10)$$ $$T(k)=\lim_{y\rightarrow -\infty,z\rightarrow
+\infty} e^{-1}(z)T(z,y)e(y).\eqno(3.11)$$ We do not reproduce the
explicit dependence on $\rho$. The functions $\Psi^\pm$ satisfy
the following integral equations:
$$\Psi^-(z)=e(z)+\int_{-\infty}^ze(z)e^{-1}(x)A^\prime(x)
\Psi^-(x)dx,\eqno(3.12)$$
$$\Psi^+(z)=e(z)-\int^{\infty}_ze(z)e^{-1}(x)A^\prime(x)
\Psi^+(x)dx,\eqno(3.13)$$ where
$$A^\prime(z)=A(z)-A_0(z)={\rho^2g_{,z}g^{-1}\over
\omega^2+\rho^2}-{\omega\over\omega^2+\rho^2}\left(\rho g_{,\rho}
g^{-1}-\pmatrix{0&0\cr0&2\cr}\right).\eqno(3.14)$$ Limit (3.11)
exists at least for $|$Im$k|>\rho$. Further, using (3.4b), one can
show that $T(k)$ does not depend on $\rho$. The basic property of
the monodromy matrix reads
$$T(z,y)=T(z,z_{2N})T(z_{2N},z_{2N-1})\ldots T(z_1,y),\;\;
z\in\Gamma_{N+1},\;y\in\Gamma_1,\eqno(3.15)$$ where $\Gamma_{N+1}$
and $\Gamma_1$ are the extreme components of the symmetry axis.
Then, from (3.15), (3.11), (3.8), (3.7) and (3.5) we get that
$$T(k)=\hat D_{N+1}\prod_{j=1,\ldots ,N} T_j\hat D_j.\eqno(3.16)$$
Here $$T_j=\pmatrix{1-{2M_j\over k-z_{2j-1}}&-4M_j\Omega_j\cr
{2L_j\over(k-z_{2j})(k-z_{2j-1})}&1+{2M_j\over k-z_{2j}}\cr},\;\;
\hat D_j=\pmatrix{1&-D_j\cr0&1\cr},$$ where the constant $D_j$ are
defined as follows: $$D_{N+1}=(\tilde Y(\infty)-\tilde
Y(z_{2N})),\;D_j=(\tilde Y(z_{2j-1})-\tilde Y(z_{2j-2})),$$
$$D_1=(\tilde Y(z_1)-\tilde Y(-\infty))$$ In (3.16), we took into
account identities (2.9) and (2.10). Notice also that $\det
T(k)=1$.

The cut of $\omega(k,z,\rho)$ is the segment that connect the points
$z+i\rho$ and $z-i\rho$. Therefore, $\Psi^\pm(k)$ are analytic
functions in $k$ as $|$Im$k|>\rho$ and
$$\Psi^-(k)=\Psi^+(k)T(k).\eqno(3.17)$$
The function $\Psi^+(k)$ ($\Psi^-(k)$) can be analytically continued
in the domains Re$k<z$ (Re$k>z$).  Using (3.5), we can see that for
$k\rightarrow\infty$,
$$\Psi^\pm(k)e^{-1}(k)\rightarrow
I\eqno(3.18)$$

Substituting $k=z+(\omega^2-\rho^2)/2\omega$ for $k$,
we pass from $\Psi^\pm(k)$ to $\Psi^\pm(\omega)$.
Then the function
$\Psi^\pm(\omega)$ become solutions to Eqs. (3.1) in the domain
$|\omega|>\rho$, while $\Psi^+(\omega)$ is analytic
in $\omega$ as Re$\omega<0, |\omega|>\rho$ and
$\Psi^-(\omega)$ is analytic as Re$\omega>0, |\omega|>\rho$.
Furthermore, it follows from (3.18) that
$$\Psi^\pm(\omega)e^{-1}(\omega)\rightarrow
I\eqno(3.19)$$
as $\omega\rightarrow\infty$.
Though $\Psi^\pm(\omega)$ are determined for
$|\omega|<\rho$ as well, they do not satisfy system
(3.1) in this domain. Therefore, our next aim is to continue
$\Psi^\pm(\omega)$ into the domain $|\omega|<\rho$ in a manner that
preserves Eqs.(3.1).

Let $\omega_1(k), \omega_2(k)$ be the roots of Eq. (3.2) and let
Re$\omega_1(k)<0$ and Re$\omega_2(k)>0$. Then the cuts of
$\omega_{1,2}(k)$ are half-lines going from points $z+i\rho$,
$z-i\rho$ to infinity in a direction that is perpendicular to the
real axis. Hence, the functions $\omega_{1,2}(k)$ are analytic for
$|$Im$k|<\rho$. Note that $\omega_1(k)=\omega(k)$ for Re$k<z$ and
$\omega_2(k)=\omega(k)$ for Re$k>z$, whence the functions
$\Psi^\pm(k)$ are continued analytically into the strip
$|$Im$k|<\rho$. Further, let $\Psi^-_1(k)$ and $\Psi^+_2(k)$ are
the solutions to Eqs. (3.12) (with $\omega_1$ substituted for
$\omega$) and (3.13) (with $\omega_2$ substituted for $\omega$),
respectively .

For $|$Im$k|<\rho$, the functions $\Psi^+(k)$ and $\Psi^-_1(k)$
($\Psi^-(k)$ and $\Psi^+_2(k)$) are the solutions of the same
differential equation (3.4a). Hence,
$$\Psi^-_1(k)=\Psi^+(k)T_1(k),\;\;\Psi^-(k)=\Psi^+_2(k)T_2(k),
\eqno(3.20)$$ where the matrices $T_{1,2}(k)$ do not depend on
$z$. Since the solutions to the integral equations (3.12) and
(3.13) automatically satisfy (3.4b) (this follows from boundary
conditions (2.12) and the fact that $e(z,\rho,k)$ is a common
solution of Eq.(3.4) with the Minkowski tensor), we conclude that
 $T_{1,2}(k)$ do not depends on $\rho$ as well.
Moreover, at $\rho\rightarrow\infty$,
$$\omega_1(k)\rightarrow
-\rho+(k-z)+O({1\over\rho}),\;\;
\omega_2(k)\rightarrow\rho+(k-z)+O({1\over\rho}).$$
Therefore, accounting for Eq.(2.12), we derive that
$$\lim_{\rho\rightarrow\infty}e_1^{-1}(k)\Psi^-_1(k)
=\lim_{\rho\rightarrow\infty}e_1^{-1}(k)\Psi^+(k)=I,$$
$$\lim_{\rho\rightarrow\infty}e_2^{-1}(k)\Psi^-(k)
=\lim_{\rho\rightarrow\infty}e_2^{-1}(k)\Psi^+_2(k)=I,
\eqno(3.21)$$
and, hence, $T_1(k)=T_2(k)=I$. Here $e_1(k)=e(\omega_1(k))$ and
$e_2(k)=e(\omega_2(k)).$
In other words, for $|$Im$k|<\rho$, we obtain
$$\Psi^-_1(k)=\Psi^+(k),\;\;\Psi^-(k)=\Psi^+_2(k).\eqno(3.22)$$
The function $\Psi^-_1(k)$ and  $\Psi^+_2(k)$ continued analytically
into the domains Re$k>z$ and Re$k<z$ respectively.
Hence, the function $\Psi^-_1(\omega)$ is analytic for
$|\omega|<\rho$, Re$\omega<0$, and the function $\Psi^+_2(\omega)$
is analytic for $|\omega|<\rho$, Re$\omega>0$. As these functions are
the solutions to Eq.(3.1) in the domain Re$\omega<0$
($\Psi^-_1$) or in the domain Re$\omega>0$ ($\Psi^+_2)$, it follows
from (3.22) that $\Psi^+(\omega)$ can be analytically continued into
the half-plane Re$\omega<0$ and remains a solution of (3.1),
and $\Psi^-(\omega)$ can be analytically continued into the
half-plane Re$\omega>0$, and also remains a solution of (3.1).

System (3.1) is invariant w. r. t. the transformation
$$\Psi(z,\rho,\omega)\rightarrow g\tilde\Psi^{-1}(z,\rho,-{\rho^2
\over\omega}),\eqno(3.23)$$ which is valid because the matrix $g$
is symmetric (the tilde in (3.23) denotes transposition).
Reduction (3.23) means that $\Psi^-(k)$ and
$g[\tilde\Psi^-_1(k)]^{-1}$ are the solutions of the compatible
pair of equations (3.4) with $\omega=\omega_1$.  However, at
$\rho\rightarrow\infty$, $$e^{-1}_2(k)ge^{-1}_1(k)\rightarrow
-I.\eqno(3.24)$$ Then from (3.21) and (3.24) we have that
$\Psi^-(k)=-g[\tilde\Psi^-_1(k)]^{-1}$ ,or, equivalently,
$$\Psi^-(\omega)=-g[\tilde\Psi^+(-{\rho^2\over\omega})]^{-1}.
\eqno(3.25)$$

The functions $\Psi^+(\omega)$ and $\Psi^-(\omega)$ satisfy the
compatible system of equations (3.1). Therefore, the combination
$[\Psi^+(\omega)]^{-1}\Psi^-(\omega)$ depends only on
$k=z+(\omega^2-\rho^2)/2\omega$ ($D_1k=D_2k=0$). Then,
by virtue of identity (3.17),
$$\Psi^-(\omega)=\Psi^+(\omega)T(k).\eqno(3.26)$$
Equations (3.26) and (3.25) show that
$$T(k)=\tilde T(k).\eqno(3.27)$$

The monodromy matrix, $T(k)$ depends on $3N+1$ parameters except
$z_k$. We treat equality (3.27) as the system of $2N+1$ nonlinear
algebraic equations for the constants $D_j, L_j$ and $\Omega_j$:
$$\sum^{N+1}_{j=1}D_j+\sum^N_{j=1}4\Omega_jM_j=0,\;
\mbox{Res}_{z_k}T_{12}(k)=\mbox{Res}_{z_k}T_{21}(k).\eqno(3.28)$$
Assume $D_j$ can be excluded from (3.28); then the remaining $N$
equations give us the connection between the angular velocities
and angular momenta. For instance, for the case of a single black
hole, it follows from (3.28) that
$$D_1=D_2=-2M_1\Omega_1,\;\;\Omega_1={L_1\over2M^2_1(m_1+M_1)}.
\eqno(3.29)$$ The expression for $\Omega_1$ in (3.29) is known; it
establishes the connection between the angular velocity and
angular momentum of the Kerr black hole.  The relationship is
expressed by $$\Omega_1={a\over(M_1+m_1)^2+a^2},\; a={L_1\over
M_1},\; M_1^2=m_1^2+a^2.$$ As in (3.29), equality (2.10) is also
taken into account here. Then the monodromy matrix reads
$$T(k)=\pmatrix{1-{2M_1\over
k-z_1}+{2M_1(M_1-m_1)\over(k-z_1)(k-z_2)}
&{2L_1\over(k-z_1)(k-z_2)}\cr{2L_1\over(k-z_1)(k-z_2)}&
1+{2M_1\over k-z_2}+{2M_1(M_1-m_1)\over(k-z_1)(k-z_2)}\cr}$$

Let us summarize the results of this section. Let boundary problem
(2.4), (2.12) have a solution. Then there exists a piecewize
analytic matrix $\Psi(\omega)$ ($\Psi(\omega)=\Psi^+(\omega),$
Re$\omega<0$ and $\Psi(\omega)=\Psi^-(\omega),$  Re$\omega>0$)
that satisfies the compatible system of linear equations (3.1),
the conjugation condition on the imaginary axis,
$$\Psi_-(\omega)=\Psi_+(\omega)T(k),\;\;{\rm Re}\omega=0
\eqno(3.30)$$ and the normalization condition at infinity,
$$\Psi(\omega)\pmatrix{1&0\cr0&{1\over\omega}\cr}\rightarrow I,
\;\omega\rightarrow\infty.\eqno(3.31)$$ In (3.30)
$\Psi_\pm(\omega)=\lim_{\epsilon\rightarrow0}
\Psi(\omega\mp\epsilon)$ $(\epsilon>0)$.

The only singularities of the matrix $T(k)$ as a function of
$\omega$ are simple poles at the points
$\omega^\pm_i=(z_i-z)\pm\sqrt{(z_i-z)^2+\rho^2}.$ Since the matrix
$$\pmatrix{1&0\cr0&\omega\cr}T(k)\pmatrix{1&0\cr0&1/\omega\cr}$$
is regular at $\omega=0$ and tends to $I$ as
$\omega\rightarrow\infty$ (which follows from explicit form of $T(k)$
and from Eqs. (3.27), (3.28)), we conclude that
$$\Psi^\pm(\omega)=\left(I+\sum_{j=1}^{2N}{A_j^\pm\over
\omega-\omega_j^\pm}\right)\pmatrix{1&0\cr0&\omega\cr},\eqno(3.32)$$
where $A_j^\pm$ do not depend on $\omega$.

Formula (3.32) demonstrate that a rational-in-$\omega$ solution to
the auxiliary linear problem corresponds to each solution with a
disconnected horizon and, hence, each such solution should be
contained in the Belinskii-Zakharov class of solutions
\cite{Belinskii-Zakharov79}.

Assume that for any $\Omega_i, z_i$
(or $L_i, z_i$)  the system of nonlinear equations (3.28)
has a unique solution. Then a unique symmetric matrix $T(k)$
corresponds to each solution of the boundary problem (2.3),
(2.4), and (2.12). However, since the solution to the Riemann problem
(3.30), (3.31) is unique and $g$ is unambiguously reconstructed by
$\Psi(\omega)$ (see 3.25), we conclude that if a solution to the
problem (2.3), (2.4), (2.12) exists, then it is unique as well.
In particular, for the case where only one black hole is present, we
obtain a new proof of the uniqueness of the Kerr solution.

The present paper was supported by the RFBR grant No. 96-01-00548.

\end{document}